# Chain Table: Protecting Table-Level Data Integrity by Digital Ledger Technology

Feng "George" Yu[1], Ryan Laird[1]

[1] Computer Science and Information Systems, Youngstown State University, Youngstown, Ohio, USA

`fyu@ysu.edu, rrlaird@student.ysu.edu`

**Abstract**

The rise of blockchain and Digital Ledger Technology (DLT) has gained wide traction. Instead of relying on a traditional centralized data authority, a blockchain system consists of digitally entangled block data shared across a distributed network. The specially designed chain data structure and its consensus mechanism protect blockchain data from being tempered by unauthorized adversaries. However, implementing a full-fledged blockchain system to protect a database can be technically cumbersome. In this work, we introduce an in-database design, named chain table, to protect data integrity without the need for a blockchain system. It features a succinct design without significant technology barriers or storage overhead. To realize rigorous data security, we also propose a set of data writing principles for the chain table. We prove that the chain table together with the data writing principles will guarantee flexible data integrity, named table-level data integrity (TDI).

## 1 Introduction

Since the invention of Bitcoin [1] in 2008, the rise of Blockchain has been widely witnessed. Different from a traditional centralized data storage model, a blockchain employs a decentralized storage model without relying on a trusted central authority. Inside a blockchain system, the data is serialized in structure units named blocks. Each block is chained with the previous block by including the cryptographic hash code of the previous block, namely the previous block hash, into its block header. Each block's hash code is generated including the previous block's hash code. This special design establishes a digitally-encrypted ledger system labeled as the Digital Ledger Technology (DLT) [2].

Blockchain systems and digital ledger systems are applied to many fields. Well-known examples include the creation of cryptocurrencies such as Bitcoin and Ethereum [3]. The blocks created in the Bitcoin are safeguarded by a cryptographic procedure named consensus algorithm. The peers in the Bitcoin network are competing with each other by hashing the current block data satisfying a target difficulty. The winner of the hashing competition will be privileged to publish a block and be awarded a certain number of Bitcoins. This consensus algorithm is commonly known as the Proof-of-Work (PoW) Different blockchain technologies employ various consensus algorithms based on their needs and application requirements, such as Proof-of-Stake (PoS), Proof-of-Authority (PoA), Practical Byzantine Fault Tolerance (PBFT), etc. [4]

One reason blockchain technology and DLT have gained wide traction is related to data integrity [5]. Many high-security level databases, such as a Social Security database, are critical and shall not allow anyone without authority to temper the data or against the ground truth. Because the blocks in a blockchain are digitally locked and witnessed in a distributed network, the blockchain can prevent most data tempering and guarantee general data integrity. Therefore, using the technology of blockchain and digital ledger systems to protect the data integrity of database systems deserves much attention.

One challenge of using blockchain to protect the data integrity of a database is that creating a blockchain or digital ledger can present technical overheads. First, it needs a distributed network to share the protected data. The data of high-level security databases are often sensitive and cannot be transmitted over an external network which can be intercepted by unauthorized users. Second, deploying a blockchain system will result in extra human and system costs.

In this work, we focus on protecting the data integrity of a database system without the need to deploy an additional blockchain system. Instead, we introduce an in-database design, named chain table, that emulates digitally chained records in a blockchain system. The chain table features a succinct design that can be easily implemented in any database without causing additional technical or storage overheads. In addition, we introduce standard writing operation principles on a chain table. With the designed structure and writing operation principles, we prove the chain table can guarantee the data integrity of its associate data table, named table-level data integrity (TDI), and withstand various threat models of data tempering.

The rest of this paper is structured as follows. Section 2 introduces fundamentals related to blockchain and data integrity. Section 3 includes the problem statement of this work. Section 4 introduces the chain table and demonstrates the table-level data integrity protected by the chain table. Section 5 includes related work. Section 6 includes the conclusion and future works.

## 2  Background

### 2.1  Blockchain Technology

A blockchain stores data in unit structures named blocks. Each block consists of two parts including the block header and the block data. A block header includes the general information of a block, for example, the index number of the block in the blockchain, and a specially designed hash code of the block named block hash. The block hash is generated by using the current block data and some additional information specified by the blockchain system. For example, in a Bitcoin blockchain, the block is hashed using the block data, the hash code of the previous block, and a random number, named nonce, running the SHA algorithm (namely SHA256) [6]. The first block in Bitcoin, namely the Genesis Block, is hashed by setting the previous block as null or empty [1].

The block data are protected from data tempering by the digitally chained design. If anyone wants to modify a record in the block data, such as manipulating the value of a cryptocurrency transaction, not only will the current block data need to be modified, but all the following blocks will also need to be modified and re-hashed. Blockchain also protects the generating of blocks by imposing a consensus mechanism, such as Proof-of-Work (PoW), which makes arbitrarily generating blocks computationally impossible [4].

## 2.2 Data Integrity

In high-security level databases, the data tables are critical, such as individuals' social security information, and cannot be exposed or manipulated by unauthorized persons or unintentional operations. Data integrity aims to prevent data being tempered or manipulated resulting in inconsistent data or against the ground truth.

In a centralized database system, many data protection mechanisms and policies can be implemented to protect data integrity. However, a challenge is how to prevent powerful adversaries, such as the database administrator, from directly manipulating a data table.

In a relational database, a data table is designed with a structure, namely the schema of the table. A common schema of a data table includes an identifier column, a timestamp column, and a data load column such as a description column. A relational table usually is designed with a primary key such as the unique identifier column to distinguish a data record from other records.

The data integrity of a relational database includes the intact of both the data records and the schema of the table. The manipulation of the schema of a relational table can be easily detected by periodically checking the historical schema and current schema of the table. However, the manipulation of a data record can be secluded and difficult to detect, especially in a large data table.

# 3 Problem Statement

In this section, we formalize the data integrity problems in a relational database. Our focus is to guarantee data safety and prevent unauthorized changes to each relational table in a database. Without loss of generality, we focus on the data integrity of each table in this discussion. The problems and solutions can be generated for the entire relational database later on.

Suppose we have a relational table named Events (or table E) inside a relational database. It mimics a critical data table where detailed operational information is stored with timestamps included. The schema of the Events table in short-hand representation is as follows.

```
Events (*opid*, *timestamp*, description)
```

where `opid` is the identifier of an event log typed in automatically increasing number; `timestamp` is a common time-related data formatted in UTC numbers used to record the exact time when the event log is written into the table; `description` is the data load where the detailed operational information is stored, for example, a text-based operation log. The primary key of this table consists of (`opid`, `timestamp`) labeled by asterisks.

## 3.1 Data Integrity Problem in a Relational Database

**Definition 1: Relational Data Integrity (RDI)**

A key problem in this work is the **relational data integrity (RDI)** which aims to persist critical relational data and prevent anyone, even high-level administrators, from tempering or altering the recorded data. The relational data to be protected by RDI includes both the data content and its relational schema. For example, once the data logs are recorded in the Events table, no one can change the `opid`, timestamp, or description in the table. Nor can anyone change the schema design of the schema of the table. In a rare circumstance, if the company needs to implement another Events table with a different schema, a newly designed Events table can be created, and import the protected historical data into the new table; however, the original Events table can be archived and protected for additional time of period to meet industrial or legal requirements. □

## 3.2 Example Data Table and Enabled Operations

Table 1 demonstrates an Events table including example data. The data included mimics a common application of a data log table. The opid column includes auto incremental primary keys of each event record ranging from 1-3. The timestamp column includes demo timestamps such as 't1' as the first timestamp. The description includes numbered demo strings for operation descriptions such as 'opt1' as the first operation description.

**Table 1: Events Table**

| opid | timestamp | description | |
|------|-----------|-------------|---|
| 1 | t1 | opt1 | Insertion 1 |
| 2 | t2 | opt2 | |
| 3 | t3 | opt3 | Insertion 2 |
| 1 | t4 | opt4 | Update 1 |

We focus on the data integrity of a critical table in this work and hence consider adopting the *append-only* data writing model on this table. This means the data is only added to the table where each addition is incorporated with a *timestamp* to assist in distinguishing multiple data operations. The append-only data model is commonly adopted in many data-critical situations which helps to assist data integrity. In a table where append-only writing is not enabled, we can add a new timestamp column to the original table and follow the append-only data writing manner described above.

To simulate general application scenarios, data operations enabled on this Events table include *insertion*, *updates,* and *deletions*. An *insertion* adds new data records to the table. An *update* changes a record from its original values to new values. Without loss of generality, we only consider an update to the description column of the Event table. For a table with multiple columns to update, we can consider their combined column values as the description to be updated. A *deletion* can be considered as a special case of update where the original data content is changed to a null value and a new timestamp is added to distinguish this operation from a normal update.

In this example of the Events table, there are four records in total including both insertions and updates. The first insertion, or insertion 1, includes one record, namely {1, t1, opt1} meaning the opid as "1", the timestamp as "t1", and the description as "opt1". Insertion 2 includes two records, namely {2, t2, opt2} and {3, t3, opt3}. The last operation is an update, or update 1, presented as an appended record {1, t4, opt4}. It changes the description of the record with opid equal to 1 to a new value "opt4" with timestamp as "4". Because of the existence of a timestamp, update 1 can be easily distinguished from insert 1. This type of append-only data writing fashion can help to keep a complete verifiable history of the data and operations in this table and can be applied to tables with a more complex structure.

**Table 2: Events Table (Actual Data)**

| opid | timestamp | description |
|------|-----------|-------------|
| 1 | t4 | opt4 |
| 2 | t2 | opt2 |
| 3 | t3 | opt3 |

Table 2 depicts the "actual" data of the Events table after all operations are serialized into a database. It can be easily noticed the first record with opid as 1 has a new description as "opt4" and a new timestamp "t4" which is greater than "t2" and "t3". These can help to identify which records are original records and which are updated records.

## 3.3 Threat Models Against the Data Integrity

Even though modern database management systems usually implement comprehensive data protection mechanisms and policies to prevent critical data from being tempered. Powerful adversaries, such as the system administrator or database administrator, can bypass security mechanisms and policies and directly manipulate any data stored in the system. External unauthorized persons can sneak into a system where the database is running and steal the administrative authorities to jeopardize the data integrity.

Example threat models that can harm the data integrity in a relational database can include the following:

(1) Direct modification to a critical data table by using regular database operations such as updates and deletions where data protection mechanisms and policies are not implemented.

(2) A powerful adversary, such as a database administrator, can bypass security mechanisms and policies to alter a data table.

(3) An authorized external user can access a database via system vulnerabilities to take control and make modifications to a data table.

All these threats will eventually create unauthorized modifications to a critical data table and cause unprecedented loss to the database system and hence its stakeholders. Conventional database systems alone cannot withstand these unconventional attacks because of their limitations.

One example is to manipulate a record such as the one with the `opid` equal to 2 and change the record from {2, t2, opt2} to {2, t2, **opt5**} without authorization or against the ground truth.

How to prevent such an adverse impact on the table is a key challenge in this work. The first goal is to detect unauthorized data manipulations. As both records have the same timestamp, it's hard to detect in a database the second record is an unauthorized manipulation. The second goal is to perform data integrity checks or verify the data of a critical table is intact or has been tempered.

## 4 Relational Data Integrity Solution

As discussed in the previous sections, a single database system cannot withstand attacks on the data integrity of critical tables. Unauthorized users or powerful adversaries can manipulate records in a critical data table without leaving any traceable evidence.

Blockchain technology-empowered modern cryptography is designed to protect data integrity. Once a record has been packed into a block of the blockchain, it is theoretically impossible to temper the record after several new blocks are created in the chain [1]. The core of blockchain technology also relies on a decentralized community following the same consensus mechanism and distributed storage parties to guarantee overall data integrity. For a database developer with limited knowledge of blockchain, there can be non-trivial learning obstacles, and it can be technically cumbersome to deploy a new blockchain system simply to protect one or two critical tables in a database.

### 4.1 Chain Table

Instead of creating a new blockchain ecosystem on top of the current database, this work introduces a new data structure along the side of a critical table with a simple structure design, named chain table, in order to protect the data integrity and prevent potential threats against the data table.

We name the original table as the **data table** and the new structure to protect the data table as **a chain table** or a **ledger table** inspired by blockchain and digital ledger technology (DLT). The chain table has a simple schema which can typically include four attributes. Based on the Events table structure, the schema of its chain table or ledger table named EventLedger is presented as follows.

$$\text{EventLedger (*lid*, hash, prevHash, update)}$$

where the `lid` is an auto-increasing number identifier of each chain record and the primary key of this table; `update` is the data record(s) in an update operation; `hash` is a hash code produced from the lid, update, and the hash code for the previous chain record, namely the `prevHash` column. Inspired by blockchain technology, the production of a hash code is defined as follows.

$$\text{hash = SHA(SHA(lid, update, prevHash)} \tag{1}$$

The hashing method in this work is the SHA algorithm, for example, SHA256, which is a one-way cryptographic safe hashing method with satisfying running performance even on large data input. To reduce the probability of producing duplicated hash codes, the hashing method for the chain table employed the double SHA (or double SHA256) procedure which is also used by major blockchains such as Bitcoin. Not only does the hash code for the current record rely on the current lid and update data, but it's also associated or chained with the previous data, namely the previous hash code or `prevHash`. This design mimics the blocks in a blockchain which are securely chained together any manipulation in any block will result in breaking the hash code chain and requires repeatedly hashing all chained data after the block.

Table 3: Chain Table (EventLedger)

| lid | hash | prevHash | update |
| --- | --- | --- | --- |
| 1 | h1 | NULL | [{1, t1, opt1}] |
| 2 | h2 | h1 | [{2, t2, opt2}, {3, t3 opt3}] |
| 3 | h3 | h2 | [{1, t4, opt4}] |

Table 3 depicts the chain table for the Events table, named the EventLedger table. Since there occurred three batches of updates on the Events table, there are three records in the EventLedger table identified by the auto-incremental ledger ID or `lid` ranging from 1 to 3. Each update includes the data records newly appended to the data table. For example, the update for `lid` equals to 1 is [{1, t1, opt1}] where the squared bracket "[ ]" denotes an array of updated records and the curly bracket "{ }" denotes one updated record in the array. In a realistic application, a developer can use any fashion of array data to present the updated data records. The hash code for the first record, namely h1, is generated using Equation (1) where the prevHash is NULL or empty.

The updates of the second chain record consists of [{2, t2, opt2}, {3, t3 opt3}] where two records have been inserted. Its current hash code, namely h2, is produced by using the current update data and the previous hash code, namely h1. The updates of the last chain record consist of [{1, t4, opt4}] where one record is updated instead of inserted. The hash code h3 is produced by hashing the current update and the previous hash code h2.

To reduce hashing overhead, instead of hashing the entire table data, the current hash code in a chain record only needs the current update data and the previous hash code. This significantly reduces the time needed to generate a hash code and increases the chain table updating time given a large database. Nonetheless, the updates recorded in the chain record can be used to reconstruct the entire data table at any time after that.

## 4.2 Chain Table Storage

To prevent powerful adversary attacks, it's recommended that the chain table be stored on an independent trusted storage that the powerful adversaries cannot directly access. This will prevent a powerful adversary from directly manipulating the chain table. Unlike a blockchain system, it doesn't need to be shared with all peers in a distributed network. For example, the trusted storage can be either on an additional server or an external storage device safeguarded by a third party.

## 4.3 Principles of Write Operations

To guarantee the integrity of the data table, we introduce the following principles of write operations on the chain table.

First, the only allowed write operation on a chain table is appending data. Each batch of data table updates will be appended at the end of the chain table. This eliminates the possibility that someone, even the high privilege adversaries, can directly manipulate the chain records. In addition, the appending-only fashion of data writes can benefit the write performance on a large chain table because writing only occurs at the end of the table and avoids random searching in a large table.

Second, the chain table only accepts one record written at one time. It means each time only one record can be appended to the chain table. Any writing operation trying to update multiple records in the chain table will be prohibited and aborted.

# 5 Data Integrity with a Chain Table

The introduced chain table guarantees the **Table-Level Data Integrity (TDI)** throughout the life cycle of a data table. It means the data integrity of the data table is protected by its chain table from its creation time. One chain table only focuses on the data integrity of its associated data table but nothing else.

There are multiple *threat models* that attempt to manipulate a critical data table. Their final goal is to modify the data table to be inconsistent with the ground truth data. With the employment of a chain table, these malicious modifications will be prevented or identified and aborted.

We illustrate the table-level data integrity using the example tables, namely Events and EventLedger. We demonstrate common data attacks on the Events table and how the EventLedger will protect the data integrity of the Events table.

*Data Attack Scenario 1 (Manipulating an intermediate record):*

An adversary wants to temper a data record but not the last record in the data table. Suppose one attempts to modify the second record {2, t2, opt2} to {2, t2, **opt5**} meaning tempering the description from "opt2" to "**opt5**". Instead of appending the new record {2, t2, opt5} in the data table, this adversary wants to impose an update directly on the data table located by opid equals 2. The tempering data update is shown in Table 3 where opt5 is against the ground truth history and highlighted.

Protected by the chain table, namely EventLedger, the adversary not only needs to modify the data table but also the chain table. To modify directly on the location where opid equals 2, one needs to modify from the record including this data record in the EventLedger table and all the records following that.

Table 5 demonstrates the updates needed to temper the chain table, namely EventLedger, needed by the attack in Scenario 1. In the second record, lid equals 2, the description of the data where opid equals 2 needs to be modified from "opt2" to "opt5". The hash code of the record with lid 2 needs to be regenerated as "h2*" since the updated data is changed. For the chain record with lid 3, its hash code needs to be regenerated from "h3" to "h3*", since its previous hash code is changed from "h2" to "h2*". If there are more records after this, they all need to be modified because each previous hash code needs to be regenerated in sequence.

Recall the writing operation principles of a chain table. The scenario 1 attack will result in modifying operations instead of appending data. In addition, instead of appending one record, multiple records need to be modified. These will result in violating the writing operation principles of a chain table and will be detected and aborted.

Table 4: Data Attack Scenario 1

| opid | timestamp | description |
|------|-----------|-------------|
| 1    | t1        | opt1        |
| 2    | t2        | *opt5*      |
| 3    | t3        | opt3        |
| 1    | t4        | opt4        |

Table 5: EventLedger Modification Needed for Attack Scenario 1

| lid | hash | prevHash | update |
|-----|------|----------|--------|
| 1   | h1   | NULL     | [{1, t1, opt1}] |
| 2   | h2*  | h1       | [{2, t2, *opt5*}, {3, t3 opt3}] |
| 3   | h3*  | h2*      | [{1, t4, opt4}] |

*Data Attack Scenario 2 (Manipulating the last record):*

In attack scenario 2, an adversary attempts to modify the last record in the Event table by changing {1, t4, opt4} to {1, t4, *opt6*}. Instead of appending this new update to the end of the Event table, the adversary wants to confidentially modify the last record in place. This attack is hard to identify in the data table because the last record was originally an update. It's hard to distinguish the attack from a normal update.

In the chain table, namely EventLedger, the adversary needs to modify the last record where the lid is 3 because its update data is modified from [{1, t4, opt4}] to [{1, t4, *opt6*}]. In addition, its hash code needs to be regenerated to "h3*" from "h3".

The scenario 2 attack will result in a modification instead of an appending operation in the chain table. This violates the writing operation principles of the chain table. Therefore, it will be detected and aborted.

In both data attack scenarios, the data tempering not only needs to modify the data table but also requires modifications in the chain table. The design of the chain table and its writing operation principles can successfully detect the threats and prevent the tempted modifications from jeopardizing the data integrity on the data table.

### 5.1.1. Verification and Data Reconstruction

The chain table can not only protect the table-level data integrity but can also be used to perform integrity verification at any time. Even if the data table is found been manipulated, the chain table can be used to reconstruct a new data table.

*Verification:*

The schema of the chain table includes the update column which includes all historical data updates to the data table. At any one time when a data integrity check is needed, the system can read the updates from the chain table to construct a verification table. If at any chain record, the verification table and the data table are not consistent, a data tempering is detected.

Table 6: Data Attack Scenario 2

| opid | timestamp | description |
|---|---|---|
| 1 | t1 | opt1 |
| 2 | t2 | opt2 |
| 3 | t3 | opt3 |
| 1 | t4 | *opt6* |

Table 7: EventLedger Modification Needed for Attack Scenario 2

| lid | hash | prevHash | update |
|---|---|---|---|
| 1 | h1 | NULL | [{1, t1, opt1}] |
| 2 | h2 | h1 | [{2, t2, opt2}, {3, t3 opt3}] |
| 3 | h3* | h2 | [{1, t4, opt6}] |

*Data Reconstruction:*

Reconstruction of a data table can be due to either a data tempering has been detected, or it's damaged from a system incident. The reconstruction process simply needs to read the historical updates from the chain table and append all updates to the recreated data table.

# 6 Related Work

Blockchain and digital ledger technology (DLT) have gained wide traction due to their successful application in cryptocurrencies [7], [8], [9]. Among them, Bitcoin [10] and Ethereum [11] are the two largest cryptocurrencies globally traded and exchanged. Their success has proved using a blockchain system can protect the data integrity in an information system. However, implementing a blockchain system into a relational database will introduce significant tech obstacles and can be prohibited by high-secure-level databases where sharing data over a distributed environment is impossible.

Much research has been initiated on closing the gap in data integrity for a relational database. Some solutions offer in-database implementation of immutable and temper-proof database structures. Amazon QLDB [12] stores data in a blockchain and can verify the data integrity at a document level. However, QLDB is a document-based database solution that doesn't enable structural relational table design. Oracle Blockchain Table [13] is fully integrated into the Oracle database. It locks the hash code of the database table inside the database and can be difficult to transfer from one database system to another. SQL Ledger [14] is integrated with Azure Immutable Blob Storage of Azure SQL to provide table-level data integrity protection. To protect an updatable data table, it requires three structures, namely a ledger table, a ledger view, and a history table, to work together which features more complex data structures. In addition, it needs to maintain a Merkle tree for each data table to monitor data tempering which creates extra storage and computation overheads.

Many database systems are also designed to store the data for a blockchain system, such as the BigChainDB [15]. They focus more on implementing a new database to realize a blockchain rather than protecting the data integrity in an existing database.

# 7 Conclusion and Future Work

In this work, we introduce an in-database design to protect the data integrity of a relational database table, named chain table. Instead of deploying a blockchain system, a chain table can be simply implemented for a critical data table instead of the entire database to protect the table-level data integrity. The chain table features a succinct schema design without additional technical overheads. We also introduced a set of writing operation principles working with the chain table. Any data tempering on the data table will be detected and aborted.

In the future, we will investigate schemes to implement the chain tables in a database with more complex designs. In addition, we will investigate how to protect the data integrity of semi-structured and unstructured data management systems such as NoSQL databases.